\documentclass[3p,times,twocolumn]{elsarticle}

\usepackage{ecrc}

\volume{00}

\firstpage{1}

\journalname{Nuclear Physics B Proceedings Supplement}

\runauth{}

\jid{nuphbp}

\jnltitlelogo{Nuclear Physics B Proceedings Supplement}

\usepackage{amssymb}

\usepackage[figuresright]{rotating}

\begin{document}

\begin{frontmatter}



\dochead{}

\title{Neutrino electromagnetic properties and magnetic moment induced transition
        of neutrino between different mass states}


\author[SINP]{A. Grigoriev}
\ead{ax.grigoriev@mail.ru}
\author[MSU]{A. Lokhov}
\ead{lokhov.alex@gmail.com}
\author[MSU,JINR]{A. Studenikin}
\ead{studenik@srd.sinp.msu.ru}
\author[MIPT]{A. Ternov}
\ead{ternov@mail.com}
\address[SINP]{Skobeltsyn Institute of Nuclear Physics, Moscow State
University, 119991 Moscow, Russia}
\address[MSU]{Department of Theoretical Physics, Moscow
State University, 119991 Moscow, Russia}
\address[JINR]{Joint Institute for Nuclear Research, 141980
Dubna, Russia}
\address[MIPT]{Department of Theoretical Physics, Moscow Institute for Physics and Technology,
141700 Dolgoprudny, Russia}

\begin{keyword}
neutrino \sep magnetic moment

\end{keyword}

\end{frontmatter}



Neutrino electromagnetic properties reveal an essential area for
search for new physics beyond the Standard Model (refer to
\cite{GiuStu09} for the recent review). The number of experiments
makes it rather natural that neutrino possesses nonzero mass as
well as magnetic and, probably, electric moments. With these
fundamental characteristics of neutrino here come to light the
possibility of multiple applications of them, particularly in
astrophysics and cosmology.

Due to nonzero diagonal and transition magnetic moments neutrino
is able to interact with photons. Considering neutrino propagation
in matter (for example, in a neutron star), a new mechanism of
electromagnetic radiation by neutrino has been proposed
\cite{LobStuPLB03} and it is called the Spin Light of neutrino
(SL$\nu$). The quantum theory of the process has been developed in
\cite{StuTerPLB05}, based on the method of exact solutions \cite{Stu06Stu08} 
of the modified Dirac equation. Neutrino interaction with matter
averages in energy difference between initial and final neutrino
states providing an opportunity to emit light. Note that the
effect critically depends on the neutrino helicity.

Here we consider the transition of neutrino between two different
mass states as a development of the theory of SL$\nu$. Though the
process of the neutrino decay in vacuum and in matter has been
studied before (for example, in \cite{PalWolf82}), in our analysis
we use neutrino wave functions, exactly accounting for the
presence of medium, while the previous investigations have
accounted for matter only in the vertex function.

Below there are the process rates for the most interesting ranges
of parameters: ultra-high, high density and quasi-vacuum cases
correspondingly\\
$\Gamma=4\mu^2\tilde{n}^3(1+\frac{3}{2}\frac{m_1^2-m_2^2}{\tilde{n}p_1}+\frac{p_1}{\tilde{n}}),
 1 \ll \frac{p_1}{m_1} \ll \frac{\tilde{n}}{p_1};$
$\Gamma=4\mu^2\tilde{n}^2p_1(1+\frac{\tilde{n}}{p_1}+\frac{m_1^2-m_2^2}
    {\tilde{n}p_1}+\frac{3}{2}\frac{m_1^2-m_2^2}{p_1^2}),
    \frac{m_1^2}{p_1^2}\ll \frac{\tilde{n}}{p_1} \ll 1;$
$\Gamma\approx\mu^2 \frac{m_1^6}{p_1^3},  \frac{\tilde{n}}{p_1}
\ll \frac{m_1}{p_1}\ll 1, m_1 \gg m_2.$ For slow and heavy initial
neutrino in vacuum we obtain the rate $\Gamma\approx
\frac{7}{24}\mu^2 {m_1^3} \sim m_1^5$, which is in agreement with
neutrino radiative decay rate \cite{PalWolf82}. Here $\mu$ is
neutrino transition magnetic moment, $\tilde{n}$ is matter density
parameter, $m$, $p$ are mass and momentum of the initial(1) and
final(2) neutrinos. For further details see
\cite{GriLokStuTer_QFEXT09}.

{\it Acknowledgements.} One of the authors (A.S.) is thankful to
Professor George Tzanakos for the invitation to attend this great
neutrino meeting.




\bibliographystyle{elsarticle-num}
\bibliography{CiteBase}







\end{document}